\documentstyle[preprint,eqsecnum,aps]{revtex}
\tighten
\begin{document}
\preprint{UAB--FT--374,\ gr--qc/951205}
\draft

\title{EXACT GRAVITATIONAL SHOCK WAVE SOLUTION OF
HIGHER ORDER THEORIES}

\author{M. CAMPANELLI}
\address{Fakult\"at f\"ur Physik der Universit\"at Konstanz\\
Postfach 5560 M 674, D-78434 Konstanz, Germany}

\author{C. O. LOUSTO}
\address{
Department of Physics, University of Utah\\
Salt Lake City, UT 84112, USA\thanks{Present Address.
Electronic Address: lousto@mail.physics.utah.edu}\\
and\\
IFAE--Grupo de F\'\i sica Te\'orica, Universidad Aut\'onoma
de Barcelona,\\ E-08193 Bellaterra (Barcelona), Spain}

\date{\today}
\maketitle

\begin{abstract}

We find an {\it exact} pp--gravitational wave solution of the
fourth order gravity field equations. Outside the (delta--like)
source this {\it not} a vacuum solution of General Relativity.
It represents the contribution of the massive, $m=(-\beta)^{-1/2}$,
spin--two field associated to the Ricci squared term in the
gravitational Lagrangian. The fourth order
terms tend to make milder the singularity of the curvature at
the point where the particle is located.
We generalize this analysis to
$D$--dimensions, extended sources, and higher than fourth
order theories. We also briefly discuss the scattering of fields
by this kind of plane gravitational waves.

\end{abstract}

\section{Introduction}

Higher order theories of gravity are the generally covariant
extension of General Relativity
when one considers in the Lagrangian nonlinear terms in the
curvature. The field equations derived by variation of this Lagrangian
contain derivatives of the metric of an
order higher than the second (except
for the Lovelock's Lagrangian constructed from
the $D$-dimensional extension
of four dimensional invariants\cite{L71}). Historically,
they have been introduced by Weyl
right after General Relativity\cite{W21}.
These theories provided a framework where to
study the unification of gravity with other fundamental fields,
and the possibility of classically avoiding
cosmological singularities. More recently,
higher order theories have been
shown to lead to inflation\cite{S80} and
dimensional reduction without the
introduction of any additional scalar
field. Nowadays, among the main motivations for their study are
their appearance (as vacuum polarization terms)
in the one-loop renormalization of
fields in curved spacetimes\cite{BD82} and in the low-energy limit of
string theory\cite{Z85}.

In four dimensions, using the Gauss--Bonnet invariant, a general
fourth-order (quadratic) covariant Lagrangian can be written as
\begin{eqnarray}
I=I_{G}+I_{m}=&&{1\over 16\pi G}\int d^4x~\sqrt{-g}~\bigg\{-2\Lambda + R
+ \alpha R^2 + \beta R_{\mu\nu}R^{\mu\nu} + 16\pi G{\cal L}_m\bigg\}~,
\label{1.1}
\end{eqnarray}
where we have not considered surface terms since they will not contribute
to the analysis of the field equations we will perform.

The field equations derived by extremizing the action $I$ are given by
\begin{eqnarray}
R_{\mu\nu}-&&{1\over 2}Rg_{\mu\nu}+\Lambda g_{\mu\nu}+\alpha
H_{\mu\nu}+\beta I_{\mu\nu}=8\pi GT_{\mu\nu} \doteq -
{16\pi G\over\sqrt{-g}}
{\partial I_m\over\partial g^{\mu\nu}}~,  \label{1.2}
\end{eqnarray}
where
\begin{equation}
H_{\mu\nu}=-2R_{;\mu\nu}+2g_{\mu\nu}\Box R-{1\over 2}g_{\mu\nu}R^2
+2R R_{\mu\nu}~, \label{1.2'}
\end{equation}
and
\begin{eqnarray}
I_{\mu\nu}=&&-2R_{\mu~;\nu\alpha}^{\alpha} +\Box R_{\mu\nu}+
{1\over 2}g_{\mu\nu}\Box R+2R_{\mu}^{~\alpha}R_{\alpha\nu}
-{1\over 2}g_{\mu\nu}R_{\alpha\beta}R^{\alpha\beta}~. \label{1.2''}
\end{eqnarray}

Upon linearization of these equations one can see that in addition
to the usual graviton field this theory contains a massive
scalar field $\phi$ related to $R$
and a massive spin-two field $\psi_{\mu\nu}$ related to $R_{\mu\nu}$
(see Ref. \cite{AEL93} for further details). Asking for both fields
to have a real mass (to recover the Newtonian limit)
leads to the, so called, no-tachyon constraints
\begin{equation}
3\alpha + \beta\geq 0~~~~,~~~~  \beta\leq 0~.\label{1.3}
\end{equation}

The value of these coupling constants can only be determined by
experiments or could be computed from a fundamental theory that
would unify gravity to the other forces in nature. It is then
expected they to be of the order of the Planck scale.

The quantum properties of these theories have extensively been
discussed\cite{BOS92}. However, several issues, such as the unitarity
problem\cite{H87} and the semiclassical instabilities\cite{PS93},
remain to be solved.
Besides, we have not even a complete understanding
of the proper solutions of fourth-order theories.
In Refs. \cite{CLA94} we
have developed a perturbative method to find solutions of the field
equations (\ref{1.2}), given a solution to the general relativistic
problem (see also Refs.\ \cite{BSZ85,PS93}). The method
essentially consists of writing
the higher derivative terms as derivatives
of the matter energy-momentum tensor
$T_{\mu\nu}$. One obtains a series
development around the General Relativity
metric in powers of the coupling
constants $\alpha$ and $\beta$.  We have thus studied the
$\alpha$ and $\beta$ corrections to
the Reissner-Nordstr\"om and straight
cosmic string metrics\cite{CL96}.

Our perturbative approach, evidently, gives no corrections when
the matter energy-momentum tensor vanishes. In other words, the method
confirms that vacuum solutions (including $\Lambda\not=0$) of General
Relativity are also solutions of higher order theories. However, {\it
the converse is in general not true}.
Higher order theories have a richer
set of vacuum solutions than General relativity. If we call this set
$\Sigma_{VHO}$ and the set of vacuum solutions of General relativity
$\Sigma_{VGR}$, its difference,
$\Sigma_{\Delta V}=\Sigma_{VHO}-\Sigma_{VGR}$,
is in general a non--empty set. From the above described inability
of the perturbative approach to find solutions
in $\Sigma_{\Delta V}$, we can deduce
that such solutions, if they exist, have to be {\it non--perturbative}
around $\alpha$ and $\beta$ equal to zero.

For black hole\cite{W84} and de Sitter\cite{KS95} solutions one can
extend the no-hair theorems valid for General Relativity to fourth order
theories in the case $\beta=0$. From where we can infer that
$\Sigma_{\Delta V}$ black holes will
have the form of the Kerr metric plus
non--analytic corrections in $\beta$ only.
The perturbative corrections in
powers of $\beta$ to the Reissner-Nordstr\"om metric have been given
in Refs. \cite{EL94,CLA94}.

In Sec.\ II we introduce the gravitational shock waves, a special case
of pp--waves solution of Einstein equations with a delta like source
term. We analyze its extension to fourth order gravity and find the
corresponding solution which is an exact proper solution of
field equations \ (\ref{1.2}).
In Sec.\ III we generalize this exact solution
to $D$ dimensional spacetime, extended sources, and theories of
order higher than the fourth. We finally, in Appendix\ A, deal with
the problem of the scattering of a scalar field by these shock
wave geometries and compute the $S$--matrix for the case of a source
obtained by boosting to the speed of light the Kerr metric.

\section{Gravitational Shock Waves}

As pointed out by 't Hooft\cite{tH87}, at energies of the order or
higher than the Planck scale the picture
of particles propagating in flat
spacetime ceases to be a good approximation. Interestingly enough the
curved metric generated by such particles has a remarkable simple form
\begin{equation}
ds^2=-du~dv+H(u,x_\perp)~du^2+dx_\perp^2~,\label{2.1}
\end{equation}
where
\begin{equation}
H(u,x_\perp)=f(x_\perp)~\delta(u)~,~~u=t-z~,~~v=t+z~.\nonumber
\end{equation}
This metric represents an impulsive gravitational wave localized in
the plane $u=0$, i.e. along the motion of the particle.
The shock wave is accompanying the particle, both traveling at
the speed of light. The profile
function $f(x_\perp)$ is the only quantity
depending on the characteristic
of the source. It is only a function of
the coordinates along to
the plane of motion, $x_\perp$. The
geometry is just flat before and after the pass
of the wave, i.e. for $u\not=0$, and is
an special case of plane fronted
with parallel rays (pp) -- waves \cite{AB95}. Geodesics are then
just straight lines outside the wave front and have a finite
discontinuity (or shift) in the $v$ coordinate as they cross $u=0$
given by (see Ref.\ \cite{DtH85} for the case of the Aichelburg--Sexl
metric, i.e. Eq.\ (\ref{2.4}) below)
\begin{equation}
\Delta v=-f(\rho_i)~,
\end{equation}
where $\rho_i$ is the coordinate distance from the origin (where the
source is located) to the point where the geodesic reaches $u=0$.

There is also an effect of spatial refraction of geodesics (see again
Ref.\ \cite{DtH85})
\begin{equation}
\cot(\theta_{in})+\cot(\theta_{re})=
{1\over2}\partial_\rho f(\rho_i)~,
\end{equation}
where $\theta_{in}$ and $\theta_{re}$ are the incident and refracted
angles respectively. Of course, locally physical measurable
quantities involve
the relative shift or refraction of nearby geodesics which involve one
further derivative with respect to $\rho$, the cylindrical coordinate
on the plane $u=0$. We shall came back to this point by the end of this
section when we will compute the Riemann tensor components which give
the geodesic deviation.

The only non--vanishing components of the Riemann tensor for metric
(\ref{2.1}) are (apart from the ones obtained by symmetry properties)
\begin{equation}
R_{iuju}=-{1\over2}\partial^2_{ij}H(u,x_\perp)~,~~i,j=\{x_\perp\}~.
\end{equation}

And the only non--vanishing components of the Ricci tensor for metric
(\ref{2.1}) are (apart from the ones obtained by symmetry properties)
\begin{equation}
R_{uu}=-{1\over2}\nabla^2_\perp H(u,x_\perp)~,
\end{equation}
where $\nabla^2_\perp$ stands for the Laplacian operator in the
$x_\perp$ space.

The curvature scalars formed out of the Ricci squared, Kretschmann,
and Curvature scalar tensors all vanish identically since,
\begin{equation}
R\equiv0~,~~R_{\alpha\beta}R^{\alpha\beta}\equiv0~,~~
R_{\alpha\beta\gamma\delta}R^{\alpha\beta\gamma\delta}\equiv0~.
\label{escalares}\end{equation}

To determine the form of the profile
function one considers a null source
represented by the following energy--momentum tensor
\begin{equation}
T_{uu}=\sigma(x_\perp)\delta(u)~,
\label{2.2}
\end{equation}
with all other components vanishing.

Next, one has to impose the field equations to the metric. Einstein
equations linearize and reduce to a Poisson equation in the
$x_\perp$ space.\cite{JEK60,DtH85}
\begin{equation}
-{1\over2}\delta(u)\nabla^2_\perp f(x_\perp)=8\pi GT_{uu}
= 8\pi G\sigma(x_\perp)\delta(u)~.
\label{2.3}
\end{equation}

The simplest source one can consider is a spinless particle,
with momentum p,
represented by $\sigma_p(x_\perp)=p\delta(x_\perp)$. Then, the
solution of Eq. (\ref{2.3}) is readily found to be (in four dimensions
for the sake of simplicity)
\begin{equation}
f_{GR}(\rho)=-8Gp\ln\left({\rho\over\rho_0}\right)~,
\label{2.4}
\end{equation}
where now the profile function only depends on the radial distance
to the origin, $\rho$, in the plane $u=0$ (as expected for a
cylindrically symmetric problem). $\rho_0$ is an integration constant
having the units of $\rho$.

Metric (\ref{2.1}) and (\ref{2.4})
was originally obtained by Aichelburg
and Sexl\cite{AS71} by boosting Schwarzschild metric along the
$z$--axis and taking simultaneously the limits $M\to0$ and $v\to1$.

The explicit form of the profile function have also been found for a
variety of sources\cite{LS90,LS91,LS92a,LS92b,BN95}
by use of the above sketched procedure.

As explained above one expects the shock wave metric to be
relevant in processes involving
energies of the order or higher than the
Planck scale. At such huge energies, higher order corrections to the
gravitational theory will also be relevant. It is of interest, then,
to study the form of the profile function, $f$,
in the fourth order theory of gravity (\ref{1.1}).

If one plugs the ansatz (\ref{2.1}) for the shock wave into the
fourth order field equations (\ref{1.2}), obtains again a
linearized equation for the profile function
\begin{eqnarray}
-{1\over2}\left[\beta\nabla^4_\perp+\nabla^2_\perp\right]H(u,x_\perp)
&=&8\pi GT_{uu}\nonumber\\
=-{1\over2}\delta(u)\left[\beta\nabla^4_\perp+\nabla^2_\perp\right]
 f(x_\perp)&=&8\pi G\sigma(x_\perp)\delta(u)~.
\label{3.1}
\end{eqnarray}
where $\nabla^4_\perp=\nabla^2_\perp\nabla^2_\perp$ is the fourth
order Laplacian in the perpendicular space.

This equation can be integrated twice to give
\begin{equation}
\left[\beta\nabla^2_\perp +1\right]H(u,x_\perp)=H_{GR}(u,x_\perp)~.
\label{3.2}
\end{equation}

We have thus reduced the original fourth order problem to a
second order one. Now, by making the decomposition
\begin{equation}
H(u,x_\perp)=H_Q(u,x_\perp)+H_{GR}(u,x_\perp)~,
\end{equation}
where the index $Q$ refers to the purely quadratic part of the
solution, and using the fact that $H_{GR}$ satisfies general
relativity equations, Eq.\ (\ref{3.2}) can be re--written as
\begin{equation}
\left[\nabla^2_\perp+{1\over\beta}\right]H_Q(u,x_\perp)=
16\pi G T_{uu}~.\label{green}
\end{equation}
We first note here that the above form of equation (but with
different constant coefficients) have been found when analyzing
shock waves on curved backgrounds in pure General Relativity
\cite{DtH85,LS89,S95,L95}. 

Let us now consider the simple example of the particle represented
by the source term  $\sigma_p(x_\perp)=p\delta(x_\perp)$, (in $D=4$).
The left hand side of Eq.\ (\ref{green}) takes the form of a
Bessel equation. Our problem can thus be solved in terms of the
Green function for a cylindrically symmetric problem. The solution
can then be expressed as
\begin{equation}
H_Q(u,x_\perp)=f_Q(\rho)\delta(u)~,~~f_Q(\rho)=
-8GpI_0(0)K_0\left({\rho\over\sqrt{-\beta}}\right)~,
\label{q}\end{equation}
where $I_0$ and $K_0$ are the modified Bessel functions
and where we have taken into account that
$\beta<0$ from the non--tachyon constraints (\ref{1.3}).

Finally, the profile function generated by a massless
uncharged particle is
\begin{equation}
f(\rho)=-8Gp\left[K_0\left({\rho\over\sqrt{-\beta}}\right)
+\ln\left({\rho\over\rho_0}\right)\right]~.
\label{3.3}
\end{equation}

Equations (\ref{2.1}) and (\ref{3.3})
thus represent the exact solution
to our problem in fourth order gravity. Note that the
coupling constant $\alpha$ does not appear in the solution. This can
be traced back to the fact that the scalar curvature $R$,
identically vanishes for
metric (\ref{2.1}), as can be readily verified.

Since in the asymptotic regime, for both small and large values of  
$\rho$,
\begin{equation}
K_0\left({\rho\over\sqrt{-\beta}}\right) = \cases{
\sim -\ln\left({\rho\over2\sqrt{-\beta}}\right),  &for small
$\rho/\sqrt{-\beta}$ \cr
\sim \left({2\rho\over\pi\sqrt{-\beta}}\right)^{-1/2}
\exp\left({-\rho\over\sqrt{-\beta}}\right), &for large
$\rho/\sqrt{-\beta}$\cr} ~, \label{4.1}
\end{equation}
we have then computed the contribution to the shock wave of the massive
(with mass $m=1/\sqrt{-\beta}$) spin--two field.
In the case of small $\rho$, the profile function assumes  the form of  
an Aichelburg--Sexl profile for a point-like source.
Also note that the dependence on $\beta$ is clearly non--perturbative,
as expected from our comments in the introduction. In fact,
for $\rho\not=0$, the profile (\ref{3.3}) gives a vacuum solution to the
fourth order theory which is {\it not} solution of General Relativity,
i.e. it is in the set $\Sigma_{\Delta V}$. For $\beta\to0$ we
recover the general relativistic metric; while  $p=0$ (no source)
gives the flat space limit\footnote{Note that here we refer to the no
source case as imposing a boundary condition on the solution such that
it represents Minkowski space; while, when we refer to a
vacuum solution for $\rho\not=0$, it is the result of imposing other
boundary conditions such that Eq.\ (\ref{3.3}) is the solution. It
is in the same sense that one refers to the Schwarzschild solution as
a vacuum solution of Einstein theory, while for $M\to0$ one finds
Minkowski space.}.

In Fig.\ \ref{figura1} we plot the resulting profile function and
compare it to the general relativistic one. The first apparent
difference is the non divergence in the origin of coordinates,
i.e. where is located the particle. Even if, as we have seen in
(Eq.\ (\ref{escalares}), for plane gravitational waves the
curvature scalars identically vanish, this does not necessarily
imply the gravitational field will be non--singular at the origin.
In fact, for the cylindrically symmetric situation we are analyzing
the only non--vanishing components of the Riemann tensor for metric
(\ref{2.1}) are (apart from the ones obtained by symmetry properties)
\begin{equation}
R_{\rho u\rho u}=-{1\over2}\partial^2_\rho H(u,\rho)~,~~
R_{\phi u\phi u}=-{\rho\over2}\partial_\rho H(u,\rho)~,
\end{equation}
and the only non--vanishing components of the Ricci tensor for metric
(\ref{2.1}) is
\begin{equation}
R_{uu}=G_{uu}=-{1\over2\rho}\partial_\rho
\left(\rho\partial_\rho H(u,\rho)\right)={1\over2\beta}H_Q(u,\rho)~.
\end{equation}

It then follows that components of the Riemann tensor like
\begin{equation}
R_{\rho u\rho}^v=\partial^2_\rho H(u,\rho)~,~~
R_{\phi uu}^\phi=-{1\over2\rho}\partial_\rho H(u,\rho)~,~~
R_{\rho uu}^\rho={1\over2}\partial^2_\rho H(u,\rho)~,
\end{equation}
will diverge logarithmically as $\rho\to0$. This can be classified
\cite{Wa84} as a ``parallelly propagated curvature singularity''.
However, we note here that the singularity is notably milder than
the $1/\rho^2$ for General Relativity.


\section{Generalizations}

Many of the present candidates to unify gravity with other
interactions consider $D$, the dimensionality of the spacetime,
bigger than four. Since Eq.\ (\ref{3.1}) can be extended
to any dimension, we find that the $D$--dimensional
generalization of Eq. (\ref{3.3}) is given by
\begin{equation}
f(\rho)=-{16\pi Gp\over\Omega_{D-3}}\left[{(-2\beta)^{2-D/2}
\over\Gamma(D/2-1)}
\left({\rho\over\sqrt{-\beta}}\right)^{2-D/2}
K_{2-D/2}\left({\rho\over\sqrt{-\beta}}\right)
+{1\over (4-D)}\left({\rho\over\rho_0}\right)^{4-D}\right]~.
\label{DP}\end{equation}
where $\Omega_{D-3}=2\pi^{D/2-1}/\Gamma(D/2-1)$
is the area unit in the $D-3$ sphere and $Gp$ carry units of
$length^{D-4}$.

In the case the source term is extended, but keeps its axial
symmetry, i.e. $\sigma(x_\perp)=\sigma(\rho)$ we have
\begin{equation}
f(\rho)=f_{GR}(\rho)+16\pi Gx^{2-D/2}\int^{x}_\infty
\left\{K_{2-D/2}(x)I_{D/2-2}(r)-I_{D/2-2}(x)
K_{2-D/2}(r)\right\}r^{D/2-1}\sigma(r)dr
\label{DE}\end{equation}
where as before the index $GR$ reffers to the solution of the problem
in Einstein theory and $x=\rho/\sqrt{-\beta}$.
When the source has a $\theta$ dependence, we can Fourier transform
it as well as the solution and it will look like Eq.\ (\ref{DE}) with
the index of the Bessel functions being $l$ now (in $D=4$ and where
$l$ refers to the corresponding Fourier mode in $\theta$).

An interesting example of extended source easily solvable is the
boosted straight string\cite{LS91}. In four dimensions
Eq.\ (\ref{3.1}) becomes an ordinary fourth order differential
equation with constant coefficients and a $\delta(y)$--like
source. The profile function takes then the form
\begin{equation}
f(y)=-8\pi Gp\left[\sqrt{-\beta}e^{-|y|/\sqrt{-\beta}}+|y|\right]
\end{equation}
where $y$ refers to the perpendicular distance to the string
measured on the plane $u=0$. One can check that this exactly
corresponds to a particle in $D=3$ dimensions from
expression\ (\ref{DP}).

One can also see the problem of quadratic theories as giving a
correction to the energy--momentum tensor (as in the semiclassical
approach to quantum gravity) and think of the problem as one
for Einstein gravity with an effective source. Then, from
Eq.\ (\ref{3.2}) we have
\begin{equation}
-{1\over2}\nabla^2_\perp f(x_\perp)=8\pi G\sigma_{eff}(x_\perp)
={1\over2\beta}(f-f_{GR})~.
\end{equation}
Hence the explicit form of $\sigma_{eff}$ in terms of integrals
of $\sigma$ can be read off of Eq.\ (\ref{DE}). For example, for
the point--like source considered in the generalization of the
Aichelburg--Sexl metric, we obtain from  Eq.\ (\ref{DP})
\begin{equation}
\sigma_{eff}(\rho)=
{2\over\Omega_{D-3}}{(-2\beta)^{1-D/2}\over\Gamma(D/2-1)}
\left({\rho\over\sqrt{-\beta}}\right)^{2-D/2}
K_{2-D/2}\left({\rho\over\sqrt{-\beta}}\right)~,
\label{extendida}\end{equation}
which represents an {\it extended} effective source.

Another kind of generalization is to consider a gravitational
Lagrangian containing terms in the curvature and its derivatives
that will generate field equations with derivatives higher than
the fourth, for example, terms like $R^3$, $RR_{\mu\nu}R^{\mu\nu}$,
$R_{\mu\nu;\lambda}R^{\mu\nu;\lambda}$, etc. When we consider
solutions to the field equations of the form Eq.\ (\ref{2.1}) there
will not be any contribution coming from the terms $R^n$ with
$n\geq2$ since $R\equiv 0$ for this metric.
Neither terms with contractions
of the curvature tensors involving powers higher than the second,
since the only non vanishing component of the Ricci tensor is
$R_{uu}$. Term involving covariant derivatives of the curvature
tensors (like $R_{\mu\nu;\lambda}R^{\mu\nu;\lambda}$) will, however,
give a contribution in the form of higher order Laplacian operators
(like $\nabla^6_\perp$ in the above example). For a theory containing
terms of up to power $N$ in the curvature tensor and its derivatives
we expect to have the following form of the field equations for
a metric of the type\ (\ref{2.1})
\begin{equation}
-{1\over2}\sum_{n=1}^N\gamma_n\nabla^{2n}_\perp H(u,x_\perp)
=-{1\over2}\left[\prod_{n=1}^N\left(\nabla^2-m^2_n\right)\right]
H(u,x_\perp)=8\pi GT_{uu}~,
\label{A3}\end{equation}
where the $m_n$'s are constants related algebraically to the
$\gamma_n$'s and can be seen as the masses of some of the particles
of the theory, for example, $m_1=0$ corresponds to the graviton,
$m_2=(-\beta)^{-1/2}$ to the massive spin 2 field of quadratic
theories, and so on.

Let $H_n$ be a solution of the equation $(\nabla^2-m^2_n)H_n=0$.
Then $H_{hom}=\sum_{n=1}^NC_nH_n$ (with $C_n$ constants to be
determined by imposing the boundary conditions of our problem),
will be a general solution to the homogeneous problem associated
to Eq.\ (\ref{A3}) as  can be easily verified. We thus construct
the general solution to the inhomogeneous problem by adding a
particular solution that can be found by use of the Wronskians
method, i.e.
\begin{equation}
H_p(u,\rho)=\sum_{n=1}^NH_n(u,\rho)
\int^\rho{W_n(u,r)\over W(u,r)}dr~.
\end{equation}

Still a further generalization of the solution can be done by
considering the wave traveling on a curved background instead
of a flat one. This can be performed following the steps Dray
and 't Hooft made in Ref.\ \cite{DtH85} for Einstein theory
and should present no further difficulties than obtaining
the coefficients in the resulting field equation at $u=0$
(generalization of Eq.\ (\ref{3.1}).)

\section{Discussion}

We have seen that a nice feature of the exact solution
Eq.\ (\ref{3.3}) is the fact that it makes milder the
curvature singularity at the location of the particle
generating the gravitational field. It is in fact one of the
historical classical motivations to introduce higher order
theories in Cosmological scenarios as we recalled in the
introduction. It is also a desirable feature when one thinks of
this quadratic theory as an effective one being the by--product
of the low--energy limit of a finite quantum theory of gravity.
We have studied an exact solution of the theory given by the
action\ (\ref{1.1}). Among the motivations for studying this
kind of theory we gave the argument of its similarity with
what one finds renormalizing to one--loop a field theory in
curved backgrounds. In this case one has also to take into
account non--local corrections in the renormalized
energy--momentum tensor. That, in fact, can be done taking
into account the results of paper\ \cite{GV91}. Let us
recall that the final result in this case would be only
valid to order $\hbar$ since the renormalized 
energy--momentum tensor was computed to that order. This point
bring us to the question of ``self--consistency'' of the
perturbative approach discussed in Ref.\ \cite{PS93}. There
it is in general considered only solutions linear in $\hbar$ (and in
the coupling constants $\alpha$ and $\beta$) for consistency
with the field equations which are considered to be only
precise to one--loop. This point of view has the advantage of
avoiding the, so called, runaway solutions; rendering thus,
Minkowski space stable. In our view this procedure is too
restrictive and precludes some well--behaved physical solutions
(see also Ref.\ \cite{FW96}).
In the case of the plane gravitational wave we have studied in
this paper we note that the solution is clearly non--perturbative
in the coupling constant $\beta$ and still the full solution
physically makes sense. The ``runaway'' or ``unphysical'' solution
($I_0$ does not appear in Eq.\ (\ref{3.3})) is naturally discarded
here. Besides, we have seen that this kind of gravitational waves
will be a solution of more general Lagrangian than (\ref{1.1})
with field equations containing derivatives of order higher than
the fourth.

Note also that had we considered the possibility of $\beta>0$
in Eq.\ (\ref{green}), we would have obtained a profile function
$f(\rho)$ proportional to the Bessel functions of the first
kind, $J_0(\rho/\sqrt{\beta})$ and $Y_0(\rho/\sqrt{\beta})$.
The problem with these solutions is that now the large $\rho$
behavior is oscillatory with a slowly decreasing amplitude [as
$(\rho/\sqrt{\beta})^{-1/2}$.] This destroys the "Newtonian"
limit of the solution, i. e. the Aichelburg--Sexl profile.

\begin{acknowledgments}
We wish to express our gratitude to an anonymous
referee whose comments lead us to determine the right value
of the constant factor appearing in Eq.\ (\ref{q}).
C.O.L. was partially supported by the Direcci\'on General de
Investigaci\'on Cient\'\i fica y T\'ecnica of the Ministerio
de Educaci\'on y Ciencia de Espa\~na, CICYT AEN 93-0474,
by the NSF grant PHY-95-07719 and by research
founds of the University of Utah. M.C. holds an
scholarship from the Deutscher Akademischer Austauschdienst.
\end{acknowledgments}

\bigskip
\noindent
{note added:} After completion of this work we found
paper\ \cite{B83} which deals with sourceless pp--waves in
higher order theories. Our results appear to be completely
compatible with those of Ref.\ \cite{B83}.

\appendix
\section{Scattering of fields by shock waves}

A further interesting application of the gravitational shock waves
was stressed by 't Hooft\ \cite{tH87}. These geometries are relevant
for ultra-high-energy scattering processes (i.e. Planck scale
scattering). Let us consider a frame of reference where particle
1 is practically at rest and can be described, in a semiclassical
approach, by a scalar field satisfying the Klein--Gordon equation
in the curved background generated by particle 2, which carries
Planckian energies in our chosen system of reference. In this case
one can compute the form of the scattering matrix\cite{dVS89}
and prove that there will not be particle production of the scalar
(or other spin) fields. For the axial symmetric case the S--matrix
takes the following form 
\begin{equation}
S({\vec p}_{\perp,1},{\vec p}_{\perp,2},\omega)={1\over 2\pi }
\int_0^\infty{J_0(\mid{\vec p}_{\perp,1}-
{\vec p}_{\perp,2}\mid\rho)~e^{i\omega f(\rho)/4}~\rho~d\rho}~.
\label{matriz}\end{equation}
In Ref.\ \cite{LS91} the S--matrix
have been computed and studied for several sources.
It is evident from the form of the
profile function\ (\ref{3.3}), that it is
difficult to find an exact analytic expression for the corresponding
S--matrix. To illustrate the use of the above expression we shall
consider a recent interesting result\ \cite{BN95} where it was
obtained the profile function for the ultrarelativistic Kerr
geometry
\begin{equation}
f(\rho)=-8Gp\Theta(\rho-a)\ln\rho +8Gp
\Theta(a-\rho)\left[-\ln\left(a+\sqrt{a^2-\rho^2}\right)
+\frac{1}{2a}\sqrt{a^2-\rho^2}\right]
\end{equation}
for the boost along the axis of symmetry and where $\Theta$ is
the step function. It is still difficult to
perform the integration in\ (\ref{matriz}). We thus approximate
the profile function to be constant inside the ``ring'' of radius $a$
and leave the exact logarithmic dependence outside it, as shown in
Fig.\ \ref{figura2}. One can easily check that this approximation
is good up to order $\rho^4$ for small $\rho$ while is also good near
$\rho=a$ where $df/d\rho$ diverges. We thus decompose the
integration interval into two pieces for the dimensionless variable
$\tilde\rho=\rho/a$, form 0 to 1 and from 1 to infinity. The final
result is given by\ \cite{GR65}
\begin{equation}
S(q,s)={(a)^{2-is}\over 2\pi q}\left[2^{-is}
e^{is/2}J_1(q)+q^{is}\left(isJ_0(q)
S_{-is,-1}(q)-J_1(q)S_{1-is,0}(q)\right)\right]
\end{equation}
where $q=(\mid{\vec p}_{\perp,1}-{\vec p}_{\perp,2}\mid)/a$,
$s=2Gp\omega$, the Mandelstam variable, and
$S_{\mu,\nu}(z)$ is a Lommel function. A relevant point to see in
the structure of the S--matrix is whether it has poles that would
eventually correspond to bound states of the system. One can see
\ \cite{LS92b} that the above expression has {\it no} poles in
the $s$ variable.
This has to do with the fact the source has an extended nature,
as opposed to the boosted Schwarzschild geometry (i.e.
Aichelburg--Sexl metric), for which there are poles\cite{tH88}
at $is$ a natural number.

The same conclusions apply to the wave (\ref{3.3}). In fact,
now the corrections due to the higher derivative theory produces
a profile function (that is what matters to the scattering of
scalar fields) which is finite at $\rho=0$, as we see in
Fig.\ \ref{figura1}. The same approximation procedure can be
also applied to the profile (\ref{3.3}). This time we approximate
the profile for, let us say, $x<3$ by a straight line given by
$f(x)\cong 0.43 x+0.116$, and the match to the general relativistic
logarithmic behavior for $x>3$.
The non--appearance of poles in the S--matrix,
again can be traced back to the fact that the $\beta$--dependent
term can be seen as an extended source (see Eq.\ (\ref{extendida})).

\begin{figure}
\caption{We compare here the profile function for the general
relativistic solution, given by the logarithmic behavior, with the
profile solution of the quadratic theory.
It is evident the regularizing effect of
the quadratic terms near the particle, located at $(\rho=0)$.
We also observe that after a few mass distances
$(m=1/\protect\sqrt{-\beta})$
from the origin the two curves become undistingishables.
Here $x=\rho/\protect\sqrt{-\beta}$, and we have substracted an
irrelevant constant from Eq.\ (\protect\ref{3.3}), and set to one
the factor $-8Gp$. We thus plot $K_0(x)+\ln(x)$.}
\label{figura1}
\end{figure}

\begin{figure}
\caption{The profile function
$\tilde f(\rho)=f(\rho)+8Gp\ln(a/\rho_0)$
for the ultrarelativistic Kerr
geometry and our approximation for $\rho\leq a$ to compute
the scattering matrix, $S$. For $\rho\geq a$ we take the exact
$(\ln\rho)$ behavior. The approximation is very good near $\rho=0$
and $\rho=a$, and allow us to get the relevant features of $S$.}
\label{figura2}
\end{figure}

\end{document}